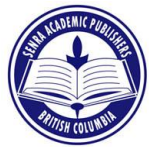

# THE COLLISION FREQUENCY IN TWO UNCONVENTIONAL SUPERCONDUCTORS

Pedro L. Contreras E
Department of Physics, University of the Andes, Mérida, Venezuela

**ABSTRACT**

The collision frequency (also known as the inverse scattering lifetime) can be self-consistently calculated from the imaginary part of the zero temperature elastic scattering cross-section in unconventional superconductors. We find these type of studies helpful to describe a hidden self-consistent damping due to incoherent fermions in two physical spaces: The Phase Space of the Nonequilibrium Statistical Mechanics, and the Configuration Space of Nonrelativistic Quantum Mechanics. The direct relation of the collision frequency with those well-known Physical Spaces is addressed in a singular way this time. Since the use of collisions for different elastic scattering regimes, is a well-developed formalism using retarded, and advanced Green functions in Many Body Physics; in order to describe our findings, we define and characterize a Reduced Phase Space for collision frequencies in the triplet strontium ruthenate compound, and the singlet doped with strontium lanthanum cuprate ceramic. Both compounds display different nodal behavior of the superconducting order parameter. In this work, their zero gap behavior is numerically scanned and used to give some illustrative examples. Finally, it is intuitively explored the "geometrical nonlocality" of the collision frequency of this type of hidden self-consistency in the Boltzmann equation, when the zero superconducting gap value drives the physics below the transition temperature, and incoherent fermions quasiparticles govern several nonequilibrium phenomena, since the macroscopic behavior remarkably changes with the strontium atomic potential strength, and the concentration inherent to both compounds.

## INTRODUCTION

This work is aimed at phenomenologically understanding the role in Nonequilibrium Statistical Mechanics (NREM) and Non Relativistic Quantum Mechanics (NRQM) (Reif, 1965; Pitaevskii *et al*., 1981; Dorfman *et al*., 2021) of one parameter used in the Boltzmann transport equation, the collision frequency of incoherent carriers (called also inverse dressed scattering fermionic lifetime $1/\tau$). This parameter enters in the exponential solution of the fermionic incoherent distribution function.

To conduct this research, two unconventional superconductors are used, the strontium ruthenathe (Maeno *et al*., 1994; Rice and Sigrist, 1995) and the doped with strontium lanthanum cuprate (Bednorz, 1986, 1988; Kastner *et al*., 1998). We categorically point out that that in both materials superconductivity is suppressed by the nonmagnetic potential of the element strontium following the well-known Larkin equation for suppression of superconductivity by nonmagnetic impurities, and its formalism (Larkin, 1965).

---

Corresponding author e-mail: pcontreras@ula.ve

It is illustrious to compare these compounds because they possess different nodal superconducting structures respect to their Fermi surfaces. These order parameters (OP) belong to different point group irreducible representations, and they have quite different transition temperatures, despite their 3D crystal structure similarities (Scalapino, 1995; Tsuei and Kirtley, 2000; Miyake and Narikiyo, 1999; Walker and Contreras, 2002; Sigrist, 2002).

Thus, the RPS helps the Phase Space to open a window in some unconventional superconductors by:

- First, defining a constrained Physical Space called "Reduced Phase Space (RPS)".
- Second, analyzing some numerical data, calculated from the self-consistent elastic cross-section by our group through several works, to prove our statements, and examples.
- Third, to build the idea that this technique provides as an instructive phenomenological approach, without entering into the microscopic superconducting mechanism.

The behavior that links the collisional frequency parameter "$\tau^{-1}$" with other spaces in Nonequilibrium Physics is sketched in Figure 1. The RPS is constructed using rationalized Planck units, and provides a NESM analysis of the incoherent (due to scattering/collisions) of fermionic quasiparticles as if the crystalline atomic strontium is seen as an external field with a collision strength, and concentration, see for example (Putzke *et al.*, 2021).

A first neighbors tight-binding approximation is used in order to keep simplicity in the physical interpretation of the acquired data, and to use the 2D nature that both materials uncover. We ought to keep in mind, that one of these superconductors is considered a triplet superconductor and has a complex 2D irrep OP, where the key issue is the breaking of the time reversal symmetry. Meanwhile in the singlet compound, time reversal symmetry is preserved owning a 1D irrep OP that lacks of time reversal symmetry breaking jumps, i.e., the specific heat and the elastic constants (alternatively the ultrasound speed in specific directions, and their polarization).

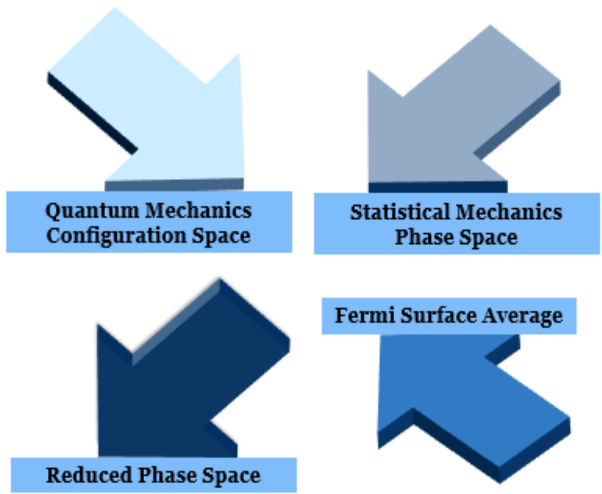

Fig. 1. The physical spaces in Nonequilibrium Statistical Mechanics (NESM) and Nonrelativistic Quantum Mechanics (NRQM) that have been successfully used to describe Solid State Physics phenomena in superconductors.

The incoherent fermionic carriers from which we collect several sets of numerical data, follow some fundamental properties:

- They obey the Fermi-Dirac statistics.
- Their inverse scattering lifetime or collisional frequency, strongly depends on the value of the magnitude of the Fermi energy and it is obtained by the Green function formalism.
- The Edwards anisotropic Fermi surface average undoubtedly depends on the zero temperature OP, and the rest of physical parameters to be mentioned.

Understand the input frequency windows needed for any of the calculation in the RPS, plays a pivotal role, since the self-consistent study of the imaginary part of the scattering cross-section is a well-established methodology developed in the 1980s, but at the same time is a numerical task, computationally very demanding (Pethick and Pines, 1986; Mineev and Samokhin, 1999).

The main purpose of this work, is to use it as an instructive computational, and pedagogical tool that helps to understand the relation between the macroscopic interpretation of different physical phenomena in these compounds such as the position of the OP nodes, when nonmagnetic disorder is added experimentally, or stoichiometric strontium is part of the crystal structure.

We clarify that we do not pretend to build a complete or even a partial theory of superconductivity in these materials. We wish only to show some of their exotic and amazing properties using this rather simplistic; but elegant, and robust picture that complement a whole set of other theories for these two materials.

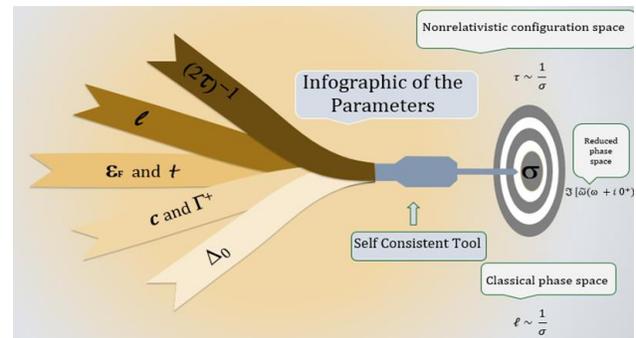

Fig. 2. The input and output physical parameters used to build the RPS.

The numerical disorder is added with the help of two parameters (Schachinger and Carbotte, 2003). The dimensionless collision parameter $c = 1/(\pi N_F U_0)$ where $U_0$ is an impurity atomic potential and $N_F$ is the density of states at the Fermi level. The other parameter is the amount of doping, or stoichiometric strength $\Gamma^+ = n_{imp}/(\pi^2 N_F)$ where $n_{imp}$ is the strontium concentration. The RPS maps a self-consistent distribution probability function always positive for the incoherent fermion carriers dressed by $U_0$.

On the other hand, NESM makes use of the parameters "$l$" and "$1/(2\tau)$". For example, for a gas of dressed Fermi quasiparticles in the case of normal metals, and alloys. A plus is that the interplay between these two parameters, makes possible to move from a complete description of a nonequilibrium state, to an abbreviated description using a single distribution function of incoherent quasiparticles (Kaganov and Lifshitz, 1989).

Elastic regimes for incoherent or dressed quasiparticles depending on the type of collision frequency in the Green function $\Im[\tilde{\omega}(\omega + i\,0^+)]$ due to nonmagnetic impurities are three (Contreras and Osorio, 2021):

- *The unitary collision regime with a maximum in $\Im[\tilde{\omega}(\omega + i\,0^+)]$ at zero frequency where holds the relation $\omega\,\tau(\tilde{\omega}) \sim 1$, and the mean free path is "l" with $l \sim a$, and is obtained from $l\,k_F \sim l\,a^{-1} \sim 1$. "$\tilde{\omega}$ – is the self-consistent frequency", "$\omega$ is the real frequency", "$k_F$ is the Fermi incoherent momentum" and "a – is the constant lattice parameter".*
- *The intermediate collision limit with a nonzero minimum in the imaginary function at the center of the imaginary function, and two maxima at finite frequencies, where hold the inequalities "$\omega \leq \tau(\tilde{\omega})^{-1}$", and "$l \geq a$".*
- *The hydrodynamic non self-consistent elastic scattering with a zero imaginary function at zero frequency i.e., "(0,0) meV" in the RPS, and two maxima in the imaginary complex function at real frequencies, holding the inequalities "$\omega \leq \tau(\tilde{\omega})^{-1}$", and "$l \geq a$".*

In this work, the physical parametrization of the RPS is made with the help of five physical quantities: the superconducting energy gap at zero temperature "$\Delta_0$ (meV)", the inverse of the scattering strength "c" (a dimensionless parameter), the concentration of nonmagnetic impurities "$\Gamma^+$(meV)", the Fermi energy of the dressed incoherent quasiparticles "$\varepsilon_F$ (meV)", and the first neighbor hoping tight-binding parameter "$t$ (meV)".

Henceforth, the tight-binding case generalizes the isotropic case (Schachinger and Carbotte, 2003; Contreras and Moreno, 2019) adding numerical anisotropy, and quantum mechanical geometrical dispersion in energy (see in Figure 2 for a graphical sketch of the parameters). The idea of using four physical input parameters self-consistently ($\Delta_0$, $\varepsilon_F$, $c$, and $\Gamma^+$) as a modeling tool in disordered HTSC was pointed out by Profs. Carbotte and Schachinger using isotropic Fermi surfaces in a series of works (Schachinger and Carbotte, 2003; Schurrer *et al.*, 1998) and all the mentions therein.

The body of this manuscript is as follows. The following section introduces the PRS. The third section analyzes the sign of the imaginary self-consistent function, and the meaning of the hidden damping, additionally links the RPS with the phase spaces of NESM and the Configuration Space in NRQM. We use numerical values obtained from the self-consistent procedure to sketch a couple of phenomenologically disordered phase diagrams for the singlet strontium doped La$_{2-x}$Sr$_x$CuO$_4$ and the triplet Sr$_2$RuO$_4$.

The fourth section calculates the values for the scattering phase-shift in these compounds using the RPS analysis. Finally, the fifth section compares briefly the mean free path, and the scattering frequency with the parameters used in the anomalous skin effect, with singular shapes in the Fermi surface for normal metals and alloys, and shortly addresses the difficult mathematical issue of nonlocality in "$l$" and "$1/\tau$". Finally, conclusions and recommendations are given.

### *The role of the "reduced phase space" linking NESP with NRQM in the two compounds*

The two dimensional self-consistent reduced phase space (RPS) for fermionic incoherent carriers is geometrically build with a pair of coordinates ($\Re(\tilde{\omega})$, $\Im(\tilde{\omega})$) in a 2D imaginary space, and it has the following properties:

- Property 1: *"The RPS in the unitary, intermedium and Born limits has two axis: the real axis $\mathcal{R}[\tilde{\omega}(\omega + i\,0^+)] = \omega$ and the imaginary axis $\Im[\tilde{\omega}(\omega + i\,0^+)]$. It serves to map a distribution complex function of dressed fermion incoherent carriers, therefore is a fermionic space".*
- Property 2: *"The relative size of the RPS changes with the variation of the symmetry of the first harmonic OP".*
- Property 3: *"Unconventional superconductors* (Mineev and Samokhin, 1999; Annett, 2004) *can be also defined as those with point or line nodes, and quasinodal regions on the Fermi surface, with an OP that has a spin paired dependence (singlet or triplet). This property allows to build self-consistently different macroscopic phases, as happen for the isotope $^3$He.*
- Property 4: *"The real part $\mathcal{R}[\tilde{\omega}(\omega + i\,0^+)]$ belongs to the x interval $\in (-\infty, +\infty)$, and the imaginary part only to the positive y axis $\in (0, +\infty)$ with the function $\Im[\tilde{\omega}(\omega + i\,0^+)] > 0$ always positive".*
- Property 5: *"The RPS resembles a space where damping is contained in the self-consistent imaginary part of the cross-section following a relationship that holds between the damping, and the imaginary par and given by the relation: $\gamma = -\Im[\tilde{\omega}(\omega + i\,0^+)]$".*
- Property 6: *"The RPS serves to calculate several superconducting, and normal state properties. Among them, the density of states (DoS)".*

The units for the input and output parameters in the RPS are the rationalized Planck units, with $\hbar = k_B = c = 1$. They are given in milielectronvolts (meV). We incorporate the tight-binding method (TB) (Harrison, 1980) into the dispersion law, the first harmonic OP, and the Edwards Fermi surface average, considering well stablished irrep. and group symmetry properties, such as the parity, and the time reversal symmetry. Due to this mixture of facts contained in several classic textbooks, the RPS opens a classical window to partially understand some macroscopic properties in the two compounds. Worthy to notice that the use of the TB enriches but at the same time complicates the calcula-tion level for the realization of the self-consistent procedure, making it more computing demanding that in the isotropic case.

### *The sign of the imaginary elastic cross-section for incoherent fermionic carriers*

The inverse of the scattering lifetime is given in normal metals, and unconventional superconductors by the following expresion $\tau^{-1}(\omega) = 2\,\Im[\tilde\omega(\omega + i\,0^+)]$ (Pethick and Pines, 1986; Mineev and Samokhin, 1999). In general, the mathematical treatment of an external constant potential "$U_0$" using the elastic scattering NRQM is a complicated subject (Landau and Lifshitz, 1981). In this work, we recall for the simplicity in the reading, that the real part is given in the RPS by the coordinate $\Re(\tilde\omega) = \omega$. The imaginary term in the RPS is represented by the function $\Im[\tilde\omega(\omega)] = (2\tau)^{-1}[\tilde\omega(\omega)]$ as the imaginary axis, with a hidden self-consistent quasi-stationary damping $\gamma = -\Im[\tilde\omega(\omega + i\,0^+)]$.

Now, let us bring to the attention some examples that address the purpose of this work. In the first instance, to describe *"self-consistent damping"* in the classical phase space for NRSM, we need the time-dependent distribution function "$f(t)$" used in the $\tau$-approximation of the Boltzmann equation, where the partial derivative respect to time refers to the collision of incoherent fermionic quasiparticles (Blatt, 1957) with $(\partial f/\partial t)_{coll} = -(f - f_0)/\tau$. If the distribution function goes rapidly to an equilibrium situation denoted by the function $f_0$, the previous expression can be approximated by

$$(\partial f/\partial t)_{coll} + 2\,\Im[\tilde\omega(\omega + i\,0^+)](f - f_0) = 0 \quad (1)$$

with a hidden self-consistent time dependent collision term, and its damping $\gamma = -\Im[\tilde\omega(\omega + i\,0^+)] = -(2\tau)^{-1}[\tilde\omega(\omega)]$. The solution of equation (1) for $f(t)$ depends on the choose for the parameters $\Delta_0$, $\varepsilon_F$, $c$, $t$, and $\Gamma^+$.

A second example, comes from the configuration space in NRQM (Kvashnikov, 2003; Davydov, 1965). If the equation for the time dependent probability density $W(t)$ is obtained with a wave function containing an extra exponential term, which describes some damping at the quasi-stationary level. This can happen for one incoherent quasiparticle inside an isotropic or anisotropic Fermi reservoir as suggested in Kvashnikov (2003). The wave function will contain quasi-stationary levels of the form $\psi_\omega(t) \sim e^{-\frac{i}{\hbar}(\epsilon - i\Gamma)t}$. For incoherent fermionic quasiparticles is known that $\Gamma/\hbar = (2\tau)^{-1}$ (Schrieffer, 1970) with a probability density $\mathcal{W}(t) = |\psi_\omega(t)|^2 = \mathcal{W}_0 e^{-2\Gamma/\hbar\,t}$ where $\mathcal{W}_0$ denotes the equilibrium case.

For $\mathcal{W}(t)$ in the configuration space (Davydov, 1965), the following equation holds $\partial \mathcal{W}(t)/\partial t = -2\Gamma/\hbar\,\mathcal{W}(t)$ (Kvashnikov, 2003). If we again look at equation (1) and rearrange this new expression as a partial differential equation with $\Gamma/\hbar = (2\tau)^{-1} = \Im[\tilde\omega(\omega + i\,0^+)]$, we obtain

$$(\partial \mathcal{W}(t)/\partial t)_{qsd} + 2\,\Im[\tilde\omega(\omega + i\,0^+)]\mathcal{W}(t) = 0 \quad (2)$$

where now *"qsd"* means quasi-stationary damping, and the partial derivative refers to quasi-stationary levels such as those that can be originated in an unconventional superconductor with strontium, from the influence on the incoherent carriers of its strontium nonmagnetic atomic potential $U_0$. Equations (1) and (2) are identical although refer to different physical processes (collision and damping). However, equation (2) resembles the $\tau$-approximation in the kinetic Boltzmann equation for NRQM.

Henceforth, we can define a hidden damping from equation (2) as being given by a coefficient $\gamma = -\Im[\tilde\omega(\omega + i\,0^+)]$ where the self-consistent mechanism will depend, how long survive the incoherent states around the atomic potential. We control the physical phases in the RPS by learning how to use properly the values of the five parameters: the hoping, the strength of the scattering, the zero superconducting gap, the Fermi energy, and the amount of disorder.

Now is clear that this analogy links the quasi-stationary probability density $W(t)$ on the NRQM configuration space (Kvashnikov, 2003) and the quasi-stationary distribution function $f(t)$ on the NESM phase space (Blatt, 1957), the last one being a classical phenomenon, the other a quantum one at the quasi-stationary level (see in Figs. 1 and 2). It can be now understood why called it a "Reduced Phase Space". The answer we find is that the *"inverse lifetime" (or collision frequency)* is the only output parameter, but the *"mean free path"* has to be given ahead by the value of the inverse strength using *"c"* of the strontium atomic potential, as an input dimensionless number. Looking at the geometrical shape of the functions numerically obtained, several phases can be predicted using the RPS.

Nonequilibrium "classical or quantum" statistical mechanics is widely related to phenomena where the damping is hidden self-consistently in the distribution probability function $f(t)$ or the quasi-stationary probability density $W(t)$ near the equilibrium, with the following coefficient:

$$\gamma[\tilde\omega(\omega + i\,0^+)] = -\Im[\tilde\omega(\omega + i\,0^+)] < 0 \quad (3)$$

Relation (3) means that the imaginary part of the elastic scattering cross-section is always defined positive and phenomenological explains the quasi-nodal points in the OP in the Miyake-Narikiyo model proposed for strontium ruthenate (Miyake and Narikiyo, 1999), where four superconducting isolate quasinodal points are symmetrically distributed in the first Brillouin zone. A second condition in the zero temperature imaginary elastic cross-section is derived from the first

$$\Im \left[ \widetilde{\omega}(\omega + i\, 0^+) \right] > 0 \qquad (4)$$

In order to validate relation (4) in the case of the two unconventional superconductors, we discuss several numerical calculations in detail.

1. We begin with Table 1 that lists a few points of the whole set of data calculated self-consistently to obtain the Miyake-Narikiyo tiny gap (Contreras *et al.*, 2022b) in the unitary collision regime with the five input values $\Delta_0 = 1.0$ meV, $\varepsilon_F = -0.4$ meV, c = 0, t = 0.4 meV and $\Gamma^+ = 0.05$ meV. As can be seen from the second column in Table 1 with values taken from the self-consistent solution of the function $\Im \left[ \widetilde{\omega}(\omega + i\, 0^+) \right]$, the numbers that represent the tiny gap are close to zero but always positive (since 1 meV = $10^{-3}$ eV), so the values of the imaginary self-consistent elastic scattering cross-section are never zero, or negative in our calculations, when the Fermi energy is negative, and far from zero value ($\varepsilon_F = -0.4$ meV). The smallest number obtained self-consistently is shadowed gray in the second column of Table 1. In (Contreras *et al.*, 2022a), we saw that for the value $\Gamma^+ = 0.01$ meV, Figure 4 has a yellow curve, where the imaginary function of the incoherent fermionic carries dies inside the superconducting frequency interval, presenting signs of an antiferromagnetic insulating phase since it becomes an insulator with no carriers, and where the zero superconducting gap cannot be recovered.

2. To complement this, some numbers for the case where $Sr_2RuO_4$ has point nodes is also showed in the third column of Table 1 (Contreras *et al.*, 2022b). The parameter for the Fermi energy is now bigger, and an order of magnitude close to the zero value ($\varepsilon_F = -0.04$ meV). To observe the numerical behavior in this second case, the other four parameters remained equal to those used in the quasinodal case, pointing that the value of $\varepsilon_F$ is the one that models the type of nodes (quasinodal or point nodes).

3. For the nodal points, we show Figure 3, where there are not small values in the imaginary part as seen in the third column of Table 1, with the minimum of the imaginary function shadowed gray, for a possible dilute but strong coalescent metallic region with the parameter $\Gamma^+ = 0.05$ meV.

At this point is good to remember that the Fermi-Dirac distribution describes the function of dressed electrons and holes on the quasi-stationary quantum energy levels (as the ones we discuss here) $\varepsilon_n$ and where n = 0,1,2… with $f_n = 1 / (e^{-\frac{\varepsilon_n - \varepsilon_f}{k_B T}} + 1)$. Therefore, it is important to recall that the Fermi energy $\varepsilon_F$ enters as a parameter in the distribution function $f_n$, and that the consequence of increasing the number of dressed incoherent fermion carriers) results in an increase of the Fermi energy (Brandt and Chudinov, 1975) as we do to obtain the point-nodes behavior in strontium ruthenate (Contreras *et al.*, 2022b). Despite strontium ruthenate continues to be part of an intense discussion with respect to its OP as expressed recentl (Curtis, M., Gradhand, M. and Annett, J. 2022. Uniaxial strain, topological band singularities and pairing symmetry changes in superconductors. DOI: https://doi.org/10.48550/arXiv.2209.00300), we find that our approach describes both situations:

- The point nodes triplet model in the unitary collision regime.
- The quasinodal triplet Miyake-Narikiyo model, also in the unitary regime.

1. Figure 3 shows the behavior of the strontium ruthenate function $\Im \left[ \widetilde{\omega}(\omega + i\, 0^+) \right]$ with parameters: $\Delta_0 = 1.0$ meV, $\varepsilon_F = -0.04$ meV, $c = 0$, $t = 0.4$ meV and $\Gamma^+ \approx (0.05\text{-}0.40)$ meV varying from dilute to optimal numbers (Contreras *et al.*, 2022b, 2022e) . From Figure 3, it can be observed for example, that only for $\Gamma^+ = 0.05$ meV there is a noticeable change in slope around the frequency value of 1.4 meV, that number is close to the $T_c$ for strontium ruthenate when crystalline bulb samples are clean. ($T_C$ is ~ 1.5 Kelvin). The other dressed curves show a smooth minimum displaced to higher RPS frequencies (Contreras *et al.*, 2022e). It we will discussed separately.

2. The case involving the HTSC $La_{2-x}Sr_xCuO_4$ is more difficult to obtain numerically because the real frequency window should suffix to locate the normal state-superconducting transition point; and in addition; we cannot extend this procedure to the antiferromagnetic insulating phase. This is due to the existence of gap values that strongly depend on doped nonmagnetic disorder (Yoshida *et al.*, 2012) and this numerical calculation is a difficult task, since it depends on the Fermi energy value. We found a real frequencies window of $\pm 120$ meV to describe properly the whole behavior of the imaginary elastic cross-section part (details of the last statement comparing the two cases to be published by the author in a separate manuscript).

3. One of the peculiarities with the compound $La_{2-x}Sr_xCuO_4$ is that $T_c$ depends on both the concentration of doped ions, and the number of $CuO_2$ layers, and makes the use of this procedure a computational endeavor where the initial frequency values are not always stable to obtain the hidden self-consistency. Similitudes and differences of the two compounds using this approach with a small frequency window is given in (Contreras *et al.*, 2022c,d). We think of a model composed by a gas of incoherent fermionic carriers (dressed by atomic strontium), that obey a Fermi liquid behavior (Walker, 2001).

4. For $La_{2-x}Sr_xCuO_4$, we show Tables 2 and 3, with some numerical results from ref. (Contreras and Osorio, 2021), for a zero superconducting gap with the value $\Delta_0 = 33.9$ meV, $\varepsilon_F = -0.4$ meV, $c = 0$, $t = 0.4$ meV and

5. $\Gamma^+ = 0.05$ meV using a linear nodal OP model (Scalapino, 1995; Tsuei and Kirtley, 2000). In Table 2, notice that the box shaded gray represents the minimum value for the imaginary self-consistent function, given in Figure 4 in orange color and represents a coalescent phase where the nonmagnetic strontium atoms stick together in a metallic small region, and the strontium atoms get the quasi-momentum transferred from the incoherent fermionic quasiparticles, but only for a very dilute amount $\Gamma^+ \approx (0.01 - 0.05)$ meV, sketched in Figure 4 with the yellow, and orange curves (Contreras and Osorio, 2021).

6. In the same Figure 4, it is observed a very small displacement of the minimum in the imaginary function $\Im[\widetilde{\omega}(\omega + i\,0^+)]$ when frequency values are increased. This behavior is more notorious in the other compound strontium ruthenate, and the varying parameter becomes the zero temperature gap as was obtained in (Contreras *et al.*, 2022e). But to notice the same behavior in the doped lanthanum, we show some values taken from Figure 4 in the third column of Table 3, where we have also shadowed some numerical fluctuations in the real frequency values in gray color, at the point where the normal-superconducting second order transition occurs, when scanning the function from dilute to optimal values of $\Gamma^+$.

7. If fermionic incoherent momentum is transferred to the strontium atoms in the crystal lattice, sticking together in a coalescing metallic state with an almost constant scattering lifetime for the whole set of real frequencies (as in Fig. 4, yellow color), it allows to adjust nonequilibrium low temperature data fairly well by using the same normal state scattering lifetime, but only if the impurity concentration is low enough with $\Gamma^+ \approx (0.01 - 0.05)$ meV.

8. This hypothesis was firstly proposed in (Schmitt-Rink *et al.*, 1986) In addition, we were able to fit ultrasound attenuation, and electronic heat transport data, for bulk crystals of strontium ruthenate at very low temperatures with a constant lifetime, by properly averaging the kinetic coeficients using TB parameters, and making use of the three sheets of the Fermi surface to account for the "*OZ-t*" hoping, thanks to what, a self-consistency procedure wasn't required (Contreras *et al.*, 2004f; Contreras, 2011).

9. In Figure 5, we give an intuitive sketch located inside the dashed blue rectangle built from Figures 3 and 4, on how looks like the superconducting part of the phase diagram in the reduced phase space for $La_{2-x}Sr_x CuO_4$, interpreting the results from the imaginary part of the zero scattering cross-section, and $\Gamma^+$ scanned from small to optimal values in the unitary collision regime (Contreras and Osorio, 2021). We observe that in the yellow region the superconducting phase is not formed, the incoherent carriers die in a short frequency interval, going to the insulating antiferromagnetic insulating phase (with not carriers).

10. Figure 6 shows a primitive sketch of the phase diagram for $Sr_2RuO_4$. Why primitive? Well, it remarkably shows to be rich in macroscopic phases but possible experimentally unstable due to the shorten of the RPS window comparing with cuprates, and the small value of the superconducting zero gap. We notice for example, phases with point nodes, quasinodes, and a phase where the incoherent carriers numerically die before reaching the zero gap value of 1 meV.

At this point, we remind that all calculations were possible thanks to the fact that we added Edwards nonmagnetic disorder. A review of the work in this direction with the original seminal references can be found in the classical textbook by Ziman (1979).

The use of the time dependence (nonequilibrium processes) in both functions $f(t)$ and $W(t)$ mentioned in the previous section is crucial to understand the physical picture underlying this approach, that comes from a well-established methodology, the elastic cross-section analysis (Pethick and Pines, 1986; Mineev and Samokhin, 1999; Schachinger *et al.*, 2003; Schmitt-Rink *et al.*, 1986) when we look at the numbers obtained in the reduced phase space for the inverse lifetime considering the unitary limit.

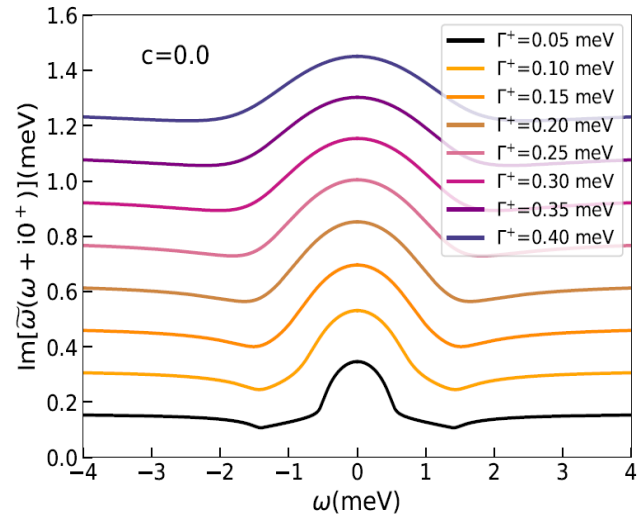

Fig. 3. The point nodes in the triplet model when the Fermi energy is very close to zero. data in Table 1 comes from the black curve calculated in (Contreras *et al.*, 2022b).

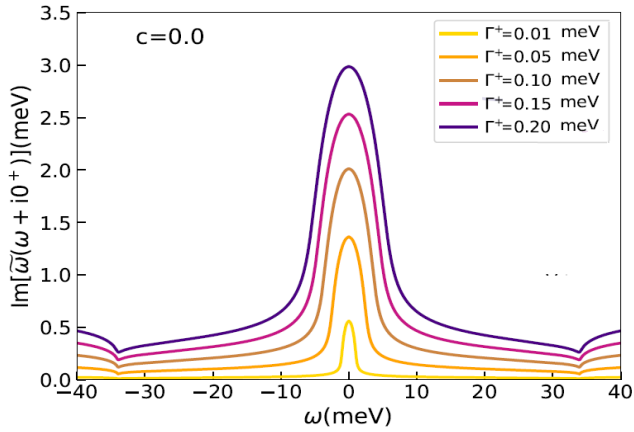

Fig. 4. The imaginary part of the elastic scattering cross-section in the unitary limit for line nodes. Data in Table 2 comes from the orange curve (Contreras and Osorio, 2021).

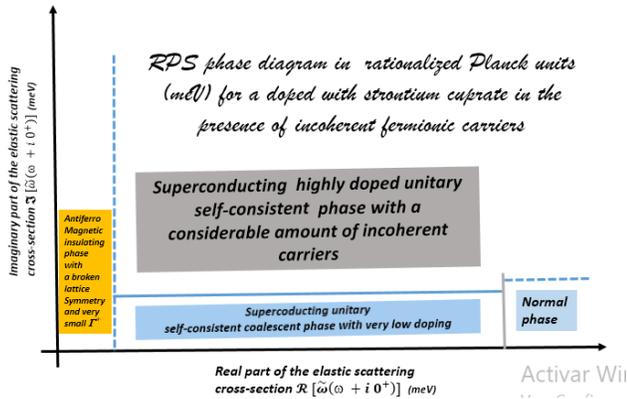

Fig. 5. The RPS from the elastic cross-section phase diagram for Lanthanum doped cuprate. For very low $\Gamma^+ = 0.01$ meV, the superconducting phase (blue part) has an almost constant lifetime.

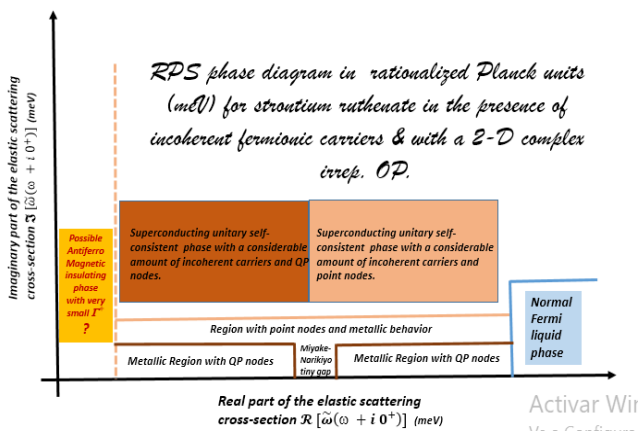

Fig. 6. The RPS from the elastic cross-section phase diagram for strontium ruthenate. It is observed that for very low $\Gamma^+ = 0.01$ MeV, the superconducting phase (survives at center and right zone, clear and dark brown, and the blue zone) or perishes (left yellow zone).

### The scattering qm phase shift $\delta_0$ and the inverse scattering strength c

In previous sections and other works, we used the RPS to numerically calculate self-consistently, and study the behavior of several families of positive collision functions depending on disorder and scattering strength, that we called in first instance "Wigner macroscopic probabilistic distributions" (Wigner, 1932; Carruthers and Zachariasen, 1983), where the energy is conserved in the three collision regimes, i.e., the unitary, the intermediate and the Born cases.

Henceforth, we calculate also the QM phase shift value, prevailing for the two compounds, and using the equation $\cot^{-1} c = \cot^{-1}(\pi N_F U_0)^{-1} = \delta_0$ (Schurrer et al., 1998). The results are obtained considering different scattering regimes. Thus, we build Table 4 that relates the inverse nonmagnetic dimensionless strength $c$ which the phase shift $\delta_0$.

As we can observe from the second column in Table 4, numerically this model shows that the HTSC unconventional cuprate superconductor $La_{2-x}Sr_xCuO_4$ has a major diversity of quantum mechanically scattering phase-shift values, then the triplet superconductor strontium ruthenate. This happens when the numerical calculation is performed for the TB values mentioned in the second section because the singlet compound can be numerically found in more regimes (strontium in this case is a doped element, and therefore is under the control of the experimentalists), i.e., the unitary, the intermediate and the hydrodynamic limits (Contreras and Osorio, 2021), meanwhile the triplet strontium ruthenate model remains most of the time in the unitary, and intermediate limits, with a stoichiometric strontium (Contreras et al., 2022a).

### Frequency dispersion relations for the anomalous skin effect versus the elastic collision frequency and non-locality

Finally, in order to gain additional credibility in the use of the RPS approach with respect to the Boltzmann kinetic equation; we conclude with a very short analysis by contrasting frequency values using the anomalous skin effect (Reuter and Sondheimer, 1948) and the examples discussed in previous sections. We first, give a brief introduction to the anomalous skin effect and after that, we assemble Table 5 to summarize section five.

### Differences between normal and anomalous skin effects

In the anomalous skin effect, the equation for the metallic impedance changes and the electronic mean free path "$l$" plays the key role. Let us, summarize the main differences between the normal and anomalous skin effect briefly to start with (Abrikosov, 1972). In the normal skin effect, the metallic impedance "$\zeta$" has the equation $\zeta = Re(\zeta) - i\, Im$ ($\zeta$) composed by equal real resistive and imaginary reactive terms, i.e., Re $\zeta$ = Im $\zeta = \sqrt{2\pi\omega/(\sigma c^2)}$. The physical behavior of an external electromagnetic field (EMF) acting on the surface of a normal metal is to penetrate it, and further decay as $\sim e^{-x/\delta}$ with an effective penetration depth of the EMF given by $\delta_{normal} = c/\sqrt{2\pi\omega\sigma}$ which does not depend on the mean free path (Abrikosov, 1972).

However, normal metals have a high conductivity "$\sigma$" when $\delta_{normal}$ is small, but at low temperatures the mean free path *"l"* becomes larger, and the Ohms Law in the local form $j = \sigma E$ cannot be applied. Thus, it is used a non-local equation (*) $j(r) = \int k_{ik}(r,r')E_k(r')\,dr'$ where the anomalous skin effect is defined by saying that the kernel of the equation (*) depends on the mean free path "$k_{ik}(r,r') \sim l$" (Kaganov et al.,1997). As a consequence, the external electric field is non-uniform, and since the normal skin effect can be derived from the kinetic equation, only if the electric field is assumed uniform, the kinetic equation in the diffusive limit for a non-equilibrium fermionic distribution function has to be solved (Kaganov et al.,1997).

The main qualitative difference between normal and anomalous skin effects in the impedance equation is given by the square root of three as a coefficient in the imaginary part of the impedance: $\zeta$ = Re($\zeta$) – $\sqrt{3}$ i Im($\zeta$). Additionally, the depth penetration has a mean free path dependence given by $\delta_{anomalous} = \sqrt[3]{c^2 l}/\sqrt{4\pi\omega a\sigma}$ with $a \sim 1$, and this dependence between the mean free path *"l"* and the anomalous penetration depth is used to plot $\zeta$.

Otherwise, normal and anomalous skin effects can be differentiate sketching (Re $\zeta$)$^{-1}$ versus $\sigma^{1/2}$, where two regions are well defined (Abrikosov, 1972). One of them, where the inverse resistive impedance has an approximate linear dependence on the square root of the conductivity, that is called the normal skin effect, and another where the resistive impedance is constant and is called the anomalous skin effect (Abrikosov, 1972).

### *Geometrical interpretation of the singular behavior in the anomalous skin effect*

To describe the anomalous skin effect in geometrical terms, we say that the anomalous skin effect happens if the fermionic quasiparticles are located in a belt of the Fermi surface with two geometrical conditions: First, $\mathbf{n}\cdot\mathbf{v}(\mathbf{p}) = 0$ where $\mathbf{n}$ is a vector normal to the metallic surface and second, $\varepsilon(p) - \varepsilon_F = 0$ (Kaganov and Contreras, 1994). The singularities in $\varepsilon(p)$ will become important for the anomalous skin effect region when the radius $\ell/\delta_{normal} \ggg 1$ and that happens when the dispersion law for fermionic incoherent quasiparticles has an equation of the type $0 \sim -\epsilon_F + |p_x|v$ +(higher order terms in momentum) (Kaganov and Contreras, 1994) which is possible if the fermionic quasiparticles obey a non-quadratic energy spectrum. In that case, $\mathbf{n}\cdot\mathbf{v}(\mathbf{p}) = 0$ is not the equation of a plane in the phase space, and the belt is not a planar curve (Kaganov and Contreras, 1994).

In this singular case the geometry of the belt is a strong function of the geometry of the Fermi surface and the direction of the vector $\mathbf{n}$. As a consequence of this, the type of connectivity changes in two different ways:
- Either a closed loop can appear or disappear in the belt (O-type singularity).
- A bridge between two loops can rupture or rejoin (X-type singularity) (Avanesyan et al., 1977).

As a consequence, non-equilibrium "kinetic" characteristics of a metal such as the anomalous skin effect, or the sound absorption have singularities of the "0" or "X" types, and the change in the shape of the belt gives "local information" about the Fermi surface. Kaganov and Contreras (1994) and Avanesyan et al. (1977) called the *p-point* responsible for this type of change in connectivity *"a critical point $p_c$"*, and showed that they are located *"along curves of parabolic points"*.

Therefore, the singularities of 0 and X types can only occur only for those metals whose Fermi surfaces have parabolic points, called also zero curvature lines (Avanesyan et al., 1977). If the metal is isotropic, then there will be an effective conductivity given by the equation $\sigma_{effective} = ia\,\sigma/(|k|\,\ell)$ with $a \sim 1$ because the number of fermionic carriers that participate in the anomalous skin effect, it is approximated by $n_{effective} \sim n\,\delta/\ell$ (Abrikosov, 1972). Thus, one can say that the effective conductivity when the Fermi surfaces are isotropic depends on the mean free path as $1/\ell$, i.e., $\sigma_{effective} \propto \sigma/(|k|\,\ell)$ and depends "only" on the characteristics of the fermionic spectrum (Abrikosov, 1972).

To finalize this brief summary, it is important to mention that the diffusive reflection in the anomalous skin effect is given by including the term $\mathbf{v}\cdot\partial f/\partial \mathbf{r}$ in the Boltzmann kinetic equation, where $\mathbf{v} = f(\mathbf{p})$ is the velocity of a fermionic carriers with $\mathbf{p}$ being a quasi-momentum in the crystal lattice (Kaganov et al.,1997) that can be omitted only in the case when the mean free path is much smaller than the distances along which the electric field changes significantly. In other words, nonlocality is neglected and the skin effect is in the normal regime, when the resistive, and the reactive parts of the impedance are equal, and the conductivity does not depend on the mean free path.

Table 1. Smallest values of the imaginary elastic scattering cross-section for the miyake-narikiyo quasi-points (Contreras *et al.*, 2022a) and the point nodes OP (Contreras *et al.*, 2022b). The parameters used are given in the main text, $\Gamma^+ = 0.05$ meV (milielectronvolts).

| $\omega = \Re(\widetilde{\omega})$ (meV) | 8.51e-001 | 8.61e-001 | 8.71e-001 | 8.81e-001 | 8.91e-001 | 9.01e-001 | 9.11e-001 | 9.21e-001 | 9.31e-001 |
|---|---|---|---|---|---|---|---|---|---|
| $(2\tau^{-1}) = \Im(\widetilde{\omega})$ Quasi-point nodes (meV) | 8.63e-008 | 3.49e-008 | 3.54e-004 | 1.59e-005 | 6.25e-007 | 2.21e-008 | 5.65e-004 | 1.87e-005 | 6.74e-007 |
| $(2\tau^{-1}) = \Im(\widetilde{\omega})$ Point nodes | 3.43e-001 | 3.43e-001 | 3.43e-001 | 3.43e-001 | 3.43e-001 | 3.43e-001 | 3.43e-001 | 3.42e-001 | 3.42e-001 |

Table 2. The smallest values of the imaginary elastic scattering cross-section for the line nodes op in the unitary limit with a zero gap $\Delta_0 = 33.94$ meV and coalescent (dilute) doping $\Gamma^+ = 0.05$ meV.

| $\omega = \Re(\widetilde{\omega})$ (meV) | 33.66 | 33.71 | 33.78 | 33.81 | 33.86 | 33.91 | 33.96 | 34.01 | 34.10 |
|---|---|---|---|---|---|---|---|---|---|
| $(2\tau^{-1}) = \Im(\widetilde{\omega})$ Line nodes (meV) | 6.06e-002 | 5.97e-002 | 5.86e-002 | 5.75e-002 | 5.61e-002 | 5.47e-002 | 5.56e-002 | 5.98e-002 | 6.33e-002 |

Table 3. The displacement in the values of the real and imaginary parts of the elastic scattering cross-section observed for the singlet linear op when the zero superconducting gap is $\Delta_0 = 33.94$ meV and doping goes from very dilute to an optimal value.

| $\Gamma^+$ (meV) | 0.01 | 0.05 | 0.10 | 0.15 | 0.20 |
|---|---|---|---|---|---|
| $\omega = \Re(\widetilde{\omega})$ (meV) | 33.950 | 33.910 | 33.900 | 33.901 | 33.801 |
| $(2\tau^{-1}) = \Im(\widetilde{\omega})$ Line nodes (meV) | 9.63e-003 | 5.47e-002 | 1.19e-001 | 1.89e-001 | 2.63e-001 |

Table 4. The calculation of the phase shift for both compounds using different regimes for elastic collisions.

| *Strontium ruthenate* | $c$ values observed from the imaginary part self-consistently (0.0 for the unitary limit and 0.4 for the intermediate scattering limit) (Contreras, Osorio and Tsuchiya, 2022b). | $\delta_0$ values for the NRQM phase in degrees calculated for the phase shift from the previous column: 90.00° for the unitary regime and 68.20° for the intermediate scattering limit. |
|---|---|---|
| *Doped strontium lanthanum cuprate* | $c$ values observed self-consistently (0.0 for the unitary limit, 0.2 for the intermediate limit, and 0.4 for the Born regime) (Contreras and Osorio, 2021). | $\delta_0$ values in the NRQM phase in degrees, found for the phase shift from the previous column: 90° for the unitary, 78.70° for the intermediate and 68.20° for the hydrodynamic limit. |

Table 5. The dispersions for the anomalous skin effect versus the two unconventional superconductors.

| Kinetical Physics Condensed Matter Phenomena. | To study in | Theoretical methods of solution | Temporal dispersion relation for the scattering lifetime | Spatial dispersion relation for the quasiparticles |
|---|---|---|---|---|
| Anomalous skin effect and surface impedance with Fermion quasiparticles. | Normal metal thin samples. | Kinetic Boltzmann eq. in the $\tau$ approximation. | $\omega \tau \ll 1$ Hydrodynamic limit | $l \gg \delta$ $\delta$ is the anomalous skin depth, the mean free path is l |
| Strange metallic phase in two unconventional superconductors | Superconducting ceramic thin samples for the doped HTSC and crystal bulb samples for the ruthenate | Numerical self-consistent equation $\Im[\widetilde{\omega}(\omega + i\, 0^+)]$ in the reduced phase space. | $\omega \tau (\widetilde{\omega}(\omega)) \sim 1$ Unitary limit | $l \sim a$ **a** is the lattice parameter |

**Anomalous singular skin effect versus incoherent frequency collisions**

A link with the previous sections arises naturally, as we seek an analogy between phase and configuration spaces, and the existence of a kernel in the integro-diferential equations that include nonlocality of the kinetic parameters *"l"* and *"τ"*. As pointed out in (Kaganov *et al*.,1997) *"to find out the explicit form of the kernel k (ik) the kinetic equation for the non-equilibrium part of the electron distribution function must be solved"*. Table 5 lists several frequency dependent dispersion relations for *"l"* and *"τ"* by comparing the two effects: the anomalous skin effect in normal metals with Fermi surfaces with parabolic points (Kaganov and Contreras, 1994) with the RPS for unconventional superconductors.

In ref. (Kaganov and Contreras, 1994), it was found theoretically the impedance in the hydrodynamic limit $\omega \tau \ll 1$ for the anomalous skin effect in thin metallic films by giving some examples using complicated 3D Fermi surfaces to average the conductivity, and the impedance. We found that the real part of the impedance strongly depends on two parameters: the mean free path, and the shape of the belts on each Fermi surface. The shape of the singular belts makes used of the topological generalized Lifshitz transitions (Kaganov *et al*.,1997).

It was noticed in (Kaganov and Contreras, 1994) that by doing an appropriate integration, two physical behaviors can be distinguished in the anomalous real part of the impedance (one of them called a singular behavior, check Figure 6 in (Kaganov and Contreras, 1994) and Table 1 in (Kaganov *et al*.,1997) for the type of singular points (Avanesyan *et al*., 1977) and the impedance dependence on the mean free path).

Hence, it was stated in (Avanesyan *et al*., 1977) that the solution for the impedance and conductivity depend sensitively on the ratio of spatial and temporal dispersions of the kinetic parameters *"l"* and *"τ"*. Therefore, we state in this work, that the solution for the imaginary function $\Im[\widetilde{\omega}(\omega + i\, 0^+)]$ (or the inverse scattering lifetime) depends sensitively on the ratio of spatial and temporal dispersions for *"l"* and *"τ"* as well, since for the analysis of the previous sections, we needed the unitary collision limit where the mean free path $l \sim a$, being a the lattice parameter, but requiring this time, a numerically self-consistent calculation of the inverse scattering lifetime $1/\tau(\widetilde{\omega})$, although it might be no obvious in this case, because the existence of the kernel is not clear.

In addition, the frequency window required in the reduced phase space for the two unconventional superconductors happens if $\omega \sim 1/\tau \sim 4\,\Delta_0$ (~ 4 meV for strontium ruthenate and ~ 120 meV for the doped with strontium lanthanum cuprate). Moreover, the tight-binding parameters $(t, \epsilon_F)$ influence strongly the Fermi surface averages, and their values are able to distinguish different OP physical phases, as it was done by comparing the singular belts in the anomalous skin effect. Therefore, the relation dispersion in the scattering lifetime that holds for the unitary collision regime in the reduced phase space might be written as stated in the introduction:

$$\omega\,\tau\,(\widetilde{\omega}(\omega)) \sim 1 \qquad (5)$$

To conclude, it is important to recall that recently, the anomalous skin effect with this type of anomalies in the Fermi surface has gained attention among the research community. Mainly for microwave applications (Torkhov *et al*., 2019; Torkhov *et al*., 2022) and in the study of non-locality phenomena in solids, as recently was theoretically and experimental realized for the compound $PdCoO_2$ Baker (2022) and Baker et al. (2022). Non-local microwave electrodynamics in ultra-pure $PdCoO_2$. DOI: https://doi.org/10.48550/arXiv.2204.14239).

**CONCLUSION AND RECOMMENDATIONS**

This work was aimed at introducing with some numerical examples the importance of two physical parameters, the mean free path and inverse scattering lifetime, both widely used in nonequilibrium Statistical Mechanics, and a brief analysis of what we have called the reduced phase space for the real and imaginary parts of the elastic scattering cross-section, using two unconventional superconductors in the unitary limit as examples, when the fermionic carriers are dressed by a nonmagnetic impurity potential, for three cases of the order parameter, the quasi-nodes, point nodes, and line nodes using a 2D anisotropic TB self-consistent parametrization with nearest neighbor hoping, showing that if stoichiometric strontium is found a very low concentration $\Gamma^+$ values, it can become numerically an insulator.

Despite, we focused our study to the unitary regime, we took into account a discussion involving three scattering regimes in the imaginary part of the elastic cross-section. We have defined a *"hidden damping parameter"* $\gamma = -\Im[\widetilde{\omega}(\omega + i\, 0^+)]$ in *"the imaginary part of the elastic scattering cross-section"*, being the last always positive, i.e., "$\Im[\widetilde{\omega}(\omega + i\, 0^+)] > 0$" obtained using a self-consistent numerical procedure. Therefore, that kind of self-consistent numerically hidden behavior might be of interest for researchers who study the Statistical Physics of non-equilibrium phenomena (classical or quantum) from a macroscopic point of view.

To conclude, several examples were analyzed in the second and previous sections. Sometimes using Tables and Figures from numerical calculations, it is also possible to give analogies between the classical phase space of the NESM and the configuration space of the NRQM, and the RPS (see Figures 1 and 2 for a graphic summary). The study of the imaginary part of the elastic cross-section not only is important for these two models of unconventional superconductors with strontium, but also is of interest for the study of fermionic and bosonic trapped gases at very low temperatures, as it has been addressed by Pitaevskii (2008). For a pedagogical review of the Ginzburg-Landau equilibrium thermodynamics analysis, complementary information is given by Contreras *et al.* (2016).

Finally, the size of the RPS windows plays a fundamental part of this problem, since it can self-consistently change, as it has been shown recently for the doped lanthanum cuprate in (Contreras *et al.*, 2023). The role of the degrees of freedom from Statistical Mechanics has been discussed in Contreras *et al.* (2023).

**Authorship contribution statement**
Pedro Contreras: Conceptualization, Methodology, Software, Investigation, Validation, writing – original draft, Supervision, Writing – review and editing.

**Declaration of competing interest**
The author declare that he has no known competing financial interests or personal relationships that could have appeared to influence the work reported in this paper.